# AN XML DRIVEN FRAMEWORK FOR TEST CONTROL AND DATA ANALYSIS

J. M. Nogiec, K. Trombly-Freytag, D. Walbridge, E. Desavouret, FNAL[*], Batavia, IL60510, USA

Abstract

An extensible component-based framework has been developed at Fermilab to promote software reuse and provide a common platform for developing a family of test and data analysis systems. The framework allows for configuring applications from components through the use of XML configurations. Extension is easily achieved through the addition of new components, and many application independent components are already provided with the framework. The core of the system has been developed in Java, which guarantees its portability and facilitates the use of object-oriented development technologies.

## 1 ARCHITECTURE

R&D environments impose specific requirements on the systems used to control tests and analyze test results. Such systems must be easily modifiable to accommodate continuously changing demands and to allow for ad hoc configured experiments and trial runs. To cope with these problems a component-based approach to developing software systems has been adopted at Fermilab's Magnet Test Facility. The multitude of existing specialized, single-purpose systems developed with help of various technologies and on different platforms is being replaced by a set of homogeneous systems built on a common framework that can run on any system that supports Java.

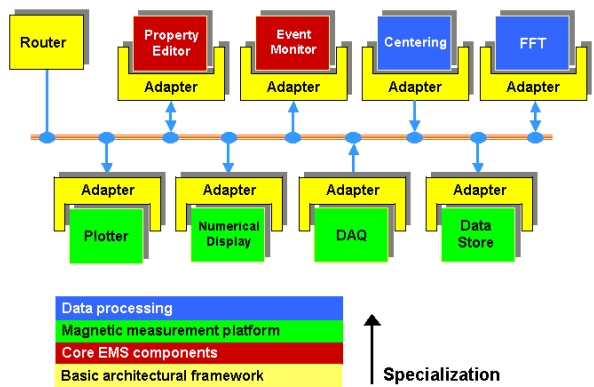

Figure 1: Architecture of the framework.

The framework consists of a set of core components and a software bus used as an inter-component communication mechanism. The bus supports multicasting and broadcasting addressing modes and its routing mechanism allows for both source routing and routing tables. The overall architecture of the framework is presented in Figure 1.

Although this framework can support various architectural patterns, it is especially suitable for a pipeline architecture, which assumes that data travels through a series of components.

## 2 APPLICATION DEVELOPMENT PROCESS

In the discussed framework, applications are assembled from components rather than being developed programmatically. Programmers develop components and domain experts assemble systems from these components. System development can be viewed as an iterative process of modifying XML setup files and running the system until the satisfactory configuration is constructed (see Fig. 2).

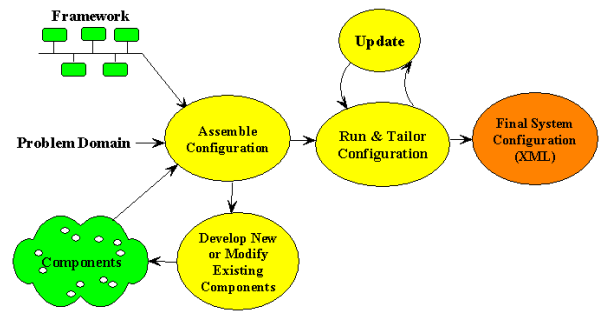

Figure 2: System Development Cycle.

## 3 XML CONFIGURATIONS

System configurations are written in a specialized XML dialect. These configurations describe components, communication patterns, and control signals. The configuration DTD (Document Type Definition) is shown in Figure 3.

```
<?xml version='1.0' encoding='UTF-8'?>
<!ELEMENT configuration (component | route | control)* >
<!ATTLIST configuration version CDATA #REQUIRED
                        title   CDATA #REQUIRED
>
<!ELEMENT component (property*)>
<!ATTLIST component id    CDATA #REQUIRED
                    class CDATA #REQUIRED
>
<!ELEMENT property EMPTY>
<!ATTLIST property name  CDATA #REQUIRED
                   value CDATA #REQUIRED
>
<!ELEMENT route EMPTY>
<!ATTLIST route type        CDATA #REQUIRED
                origin      CDATA #REQUIRED
                destination CDATA #REQUIRED
>
<!ELEMENT control EMPTY>
<!ATTLIST control signal      CDATA #REQUIRED
                  destination CDATA #REQUIRED
>
```

Figure 3: Configuration DTD.

## 3.1 Component Definition

In the presented system, components are specialized JavaBeans that implement one or more of the interfaces defined in the framework. In its XML specification (see Fig. 4) the component is identified by its *id* and is declared as an object of a given *class*. In the JavaBeans model, properties expose all the data that should be externally visible and accessible. The XML specification of the property contains its *name* and *value*. Properties allow the system assembler to tailor a component to the specific task at hand.

```
<route type="Data" origin="Producer"
                   destination="Consumer"/>
<route type="Exception" origin="*"
                   destination="ErrorLog"/>
<route type="Debug" origin="*"
                   destination="DebugWindow"/>
```

Figure 4: Example of a component definition.

## 3.2 Route Definition

Components communicate through exchange of events. A sender component is connected with its recipients via routes, which denote data paths between components (see Fig. 5). Communication patterns describe communication links between components, with separate links for data events, debug events, exception events, and control events.

```
<component id="Chart"  class="ems.core.graph.Chart">
    <property name="XPosition" value="0"/>
    <property name="YPosition" value="200"/>
    <property name="title" value="Plot Display"/>
</component>
```

Figure 5: Route definitions.

## 3.3 Controls Definition

Control signals (control events) are sent to components to request a specific action. In response, the component performs the requested action and updates its state. There exists a set of standard control signals (*init, start, stop, and exit*) and this can be extended by any number of user-defined signals. Similarly, a standard set of states is defined, which can also be extended by adding user-defined states. Example control signals to be sent to all components are shown in Figure 6.

```
<control signal="init"  destination="!"/>
<control signal="start" destination="!"/>
```

Figure 6: Example of control signals.

In configurations, control signals are used to initialize components and put them in a right initial state. This may include initializing devices, connecting to databases, setting up screens, or creating new log files.

```
<!DOCTYPE configuration SYSTEM "ems.dtd" [
  <!ENTITY coreComponents SYSTEM "core.xml">
  <!ENTITY dataComponents SYSTEM "data.xml">
]>
```

Figure 7: Example of a hierarchical setup file.

## 3.4 Hierarchy of Setup Files

As a system is constructed using more and more components, setup files could become very complex, leading to problems in system debugging and setup file reuse. Allowing a hierarchy of setup files to be created and then referenced from a single system setup file solves this problem. Each of the "leaf" files can contain XML statements for setting up a single component or a subsystem, with the root file merely including leaf files and routing between these subsystems. Figure 7 shows an example of a root setup file that includes two subsystem setup files: one that configures core system components, and the other that sets up data processing components.

## 4 TAILORING

As has been already described, component properties can be defined in XML configuration files. In addition, they can be modified at runtime, both programmatically and interactively. A component can update another component's property by sending a property event to it. The property event contains a collection of property names and their new values.

The user can also interactively examine and modify properties of any selected component at runtime. This is accomplished with help of the property editor component (see Fig. 8).

This modification of the behavior and presentation layer of an application at runtime is frequently referred to as tailoring.

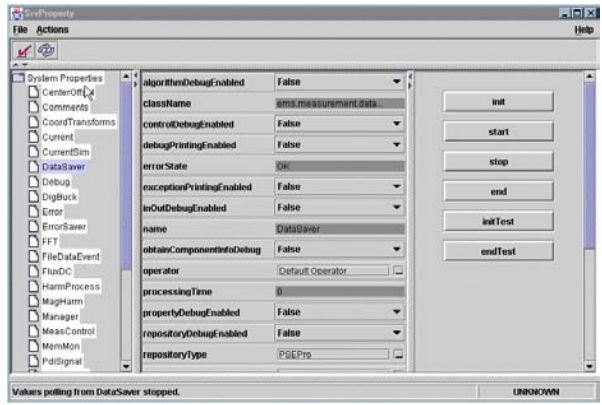

Figure 8: Property editor.

## 5 MONITORING AND DEBUGGING

In order to assist in the application development process, the system has built-in features that help in debugging and monitoring the application. Debugging of internal state and behavior of any component can be switched on or off in its XML setup file or at runtime. The amount and type of debug information is selected by enabling specific categories of this information, such as IO, algorithm, state, control, etc. In response to these settings, components generate debug events that are delivered to all the components defined in the XML configuration as recipients of debug information, e.g., display or logging components. Debug and exception information can be saved together with data in a database or/and in log files.

Special core components have also been developed to assist in monitoring of system activity at runtime. They include an event traffic monitor, a memory monitor (see Fig. 9), a processing time monitor, and a debug display.

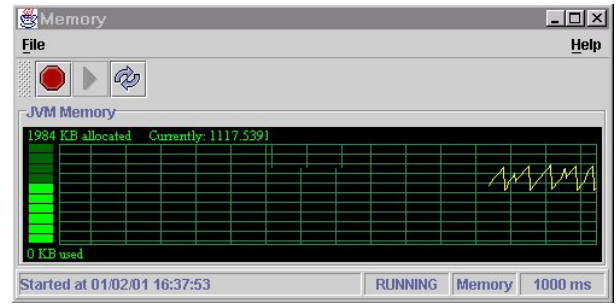

Figure 9: Memory monitor.

## 6 SUMMARY

The presented framework can support various architectures and provides a sophisticated inter-component communication mechanism. Some of its features include: support of multiple communication patterns, various built-in routing mechanisms, and provisions for distributed configurations. It is also highly tailorable, with systems constructed by selecting components and defining their properties in an XML configuration and modified dynamically by altering component properties during runtime. Specialized mechanisms are built into the framework that provide for easy troubleshooting, maintenance, and debugging. On-line monitoring of resources such as time and memory usage is also included.

The presented framework has been already successfully used to develop a system to measure the field quality of accelerator magnets [1].